\documentclass[aps,prd,secnumarabic,nobibnotes,twocolumn,superscriptaddress]{revtex4-2}
\usepackage{amssymb}
\usepackage{amsmath}
\usepackage{appendix}
\usepackage{times}
\usepackage{graphicx}
\usepackage{subeqnarray}
\usepackage{bbold}
\usepackage{booktabs}
\usepackage{enumerate}
\usepackage{mathrsfs}
\usepackage{graphics}

\usepackage{color}
\usepackage{bm}
\usepackage{multirow}

\makeatother
\makeatletter
\newcommand{\Rmnum}[1]{\expandafter\@slowromancap\romannumeral #1@} 
\makeatother

\makeatletter
\newsavebox{\@brx}
\newcommand{\llangle}[1][]{\savebox{\@brx}{\(\m@th{#1\langle}\)}%
  \mathopen{\copy\@brx\kern-0.5\wd\@brx\usebox{\@brx}}}
\newcommand{\rrangle}[1][]{\savebox{\@brx}{\(\m@th{#1\rangle}\)}%
  \mathclose{\copy\@brx\kern-0.5\wd\@brx\usebox{\@brx}}}

\begin{document}

\title{Ferroelastic Reversal Magnon Thermal Hall Effect in a Fully Compensated Ferrimagnet}

\author{Yingxi Bai}
\author{Zihan Chen} 
\affiliation
{Department of Physics and Materials, The Hong Kong Polytechnic University, Kowloon, Hong Kong SAR, China}
\author{Chengwang Niu}
\email{c.niu@sdu.edu.cn}
\affiliation
{School of Physics, State Key Laboratory of Crystal Materials, Shandong University, 250100 Jinan, China}
\author{Ming Yang}
\email{kevin.m.yang@polyu.edu.hk}
\affiliation
{Department of Physics and Materials, The Hong Kong Polytechnic University, Kowloon, Hong Kong SAR,  China}
\affiliation
{Research Centre for Nanoscience and Nanotechnology, The Hong Kong Polytechnic University, Kowloon, Hong Kong SAR, China}

\begin{abstract}
Magnons offer a charge-neutral channel for spin and heat transport, yet reversible and nonvolatile control of their thermal Hall response  poses a fundamental challenge in systems with zero net magnetization. Here, we establish ferroelastic switching as an effective route to manipulate magnonic transverse responses in fully compensated ferrimagnets. A minimal Heisenberg-Dzyaloshinskii-Moriya model reveals that ferroelastic switching can re-orientate the lattice anisotropy, which reconstructs the chirality-resolved magnon bands and Berry curvature distribution. Consequently, the two orthogonal fully compensated ferrimagnetic variants related by $[C_2 || C_{4z}]$ exhibit magnon thermal Hall responses with opposite signs, whereas the response vanishes in the intermediate altermagnetic phase. First-principles calculations further identify monolayer V$_2$SSe as a promising material candidate for realizing the proposed ferroelastic control of transverse magnon transport. Remarkably, ferroelastic switching drives a dual magnonic-electronic band reconstruction, enabling nonvolatile reversal of the magnon thermal Hall effect while producing anisotropic orbital Hall responses and distinct magneto-optical spectra as complementary transport and optical fingerprints. Our findings establish ferroelasticity as an effective route to nonvolatile magnon control and multichannel state identification in zero net magnetization systems.

\end{abstract}

\maketitle
\date{\today}

\textcolor{blue}{\textit{Introduction}}---Magnons, the bosonic quasiparticles of collective spin excitations, can carry spin angular momentum and heat without accompanying charge transport and the associated Joule heating, offering a promising platform for low-dissipation information transport~\cite{matsumoto2011theoretical,chumak2015magnon,cornelissen2015long,murakami2017thermal,mcclarty2022topological,zhang2024thermal}. The experimental observation of the magnon thermal Hall effect (MTHE) in the pyrochlore ferromagnetic (FM) insulator Lu$_2$V$_2$O$_7$ revealed a transverse heat current carried by magnons and stimulated the extension of Berry phase and band topology concepts to bosonic quasiparticles~\cite{onose2010observation}. The MTHE has since been extensively investigated in ferromagnetic systems, where broken time-reversal symmetry $\mathcal{T}$ permits net magnon Berry curvature and a finite transverse thermal response~\cite{zhang2013topological,chisnell2015topological,owerre2016first,chen2018topological,pershoguba2018dirac,nikitin2022thermal,schneeloch2022gapless,bai2024,bai2025}. 
Antiferromagnets (AFMs), despite their negligible stray fields, robustness against magnetic perturbations, and terahertz spin dynamics~\cite{jungwirth2018multiple,baltz2018antiferromagnetic,jungwirth2016antiferromagnetic}, commonly exhibit symmetry-enforced compensation. In conventional collinear AFMs, combined $\mathcal{PT}$ symmetry relates opposite-chirality magnon branches with opposite Berry curvatures, cancelling their transverse thermal contributions~\cite{zyuzin2016magnon,cheng2016spin,diaz2019topological,liang2024observation}.

More recently, unconventional magnets, particularly altermagnets (AMs) and fully compensated ferrimagnets (fFIMs), have broadened the scope of antiferromagnetic spintronics beyond conventional $\mathcal{PT}$-symmetric AFMs~\cite{vsmejkal2022emerging,bai2024altermagnetism,PhysRevLett.134.116703,song2025altermagnets,jungwirth2026symmetry,chen2025unconventional}. Their distinct spin-space group structures allow vanishing net magnetization to coexist with momentum-alternating and nonrelativistic spin splitting, respectively, thereby enabling spin-polarized currents and unconventional transverse transport~\cite{sun2025proposing,bai2025anomalous,yang2026altermagnet}. Coupling these magnetic phases to switchable ferroic orders further provides a versatile route to reversible and nonvolatile control of their spin-dependent properties. For example, the coexistence of ferroelectricity and altermagnetism enables electric-field control of nonrelativistic spin splitting and anomalous Hall responses~\cite{gu2025ferroelectric,duan2025antiferroelectric,zhu2025sliding}. Ferroelastic switching likewise enables stress-driven reconstruction of spin-split electronic bands and the associated spin-space group symmetry in fFIMs~\cite{peng2025ferroelastic,huang2025spin,che2026symmetry}. The demonstrated ferroic control of electronic states naturally motivates the exploration of analogous control over magnon transport~\cite{wang2025two,bai2026}. However, in AFM magnon systems, a finite MTHE has so far been reported only in a limited number of cases where the cancellation between opposite-chirality magnons is lifted, for instance, by inequivalent magnetic sublattices or external fields~\cite{neumann2022thermal,hoyer2025spontaneous,bai2026emergent,du2026odd}. Beyond realizing a finite response, reversible and nonvolatile switching of its sign while preserving zero net magnetization remains largely unexplored.

In this Letter, we demonstrate a ferroelastically tunable platform for reversible and nonvolatile control of transverse magnonic and electronic transport in unconventional magnets. The oppositely polarized sublattices are not connected by any spin space-group operation within each fFIM phase, thereby lifting the symmetry constraint between magnon branches of opposite chirality and allowing a finite MTHE. Nevertheless, the two orthogonal fFIM variants are mapped onto one another by $[C_2 || C_{4z}]$. Ferroelastic switching  therefore rotates the anisotropic magnon spectrum and Berry curvature distribution by $90^\circ$, exchanges the magnon chirality, and reverses the MTHE polarity while preserving zero net magnetization. To substantiate this symmetry-based mechanism, we develop a minimal Heisenberg-Dzyaloshinskii-Moriya (DM) model and examine V$_2$SSe monolayer through first-principles calculations and spin space group analysis. The material calculations validate the predicted ferroelastic reversal of transverse magnon heat transport and further reveal anisotropic orbital Hall responses together with phase-dependent Kerr and Faraday spectra. 
These results reveal a coupled magnonic-electronic response to ferroelasticity, integrating nonvolatile control of transverse magnon transport with multichannel identification of compensated magnetic states.

\begin{figure}
\centering
\includegraphics[width=8.5cm]{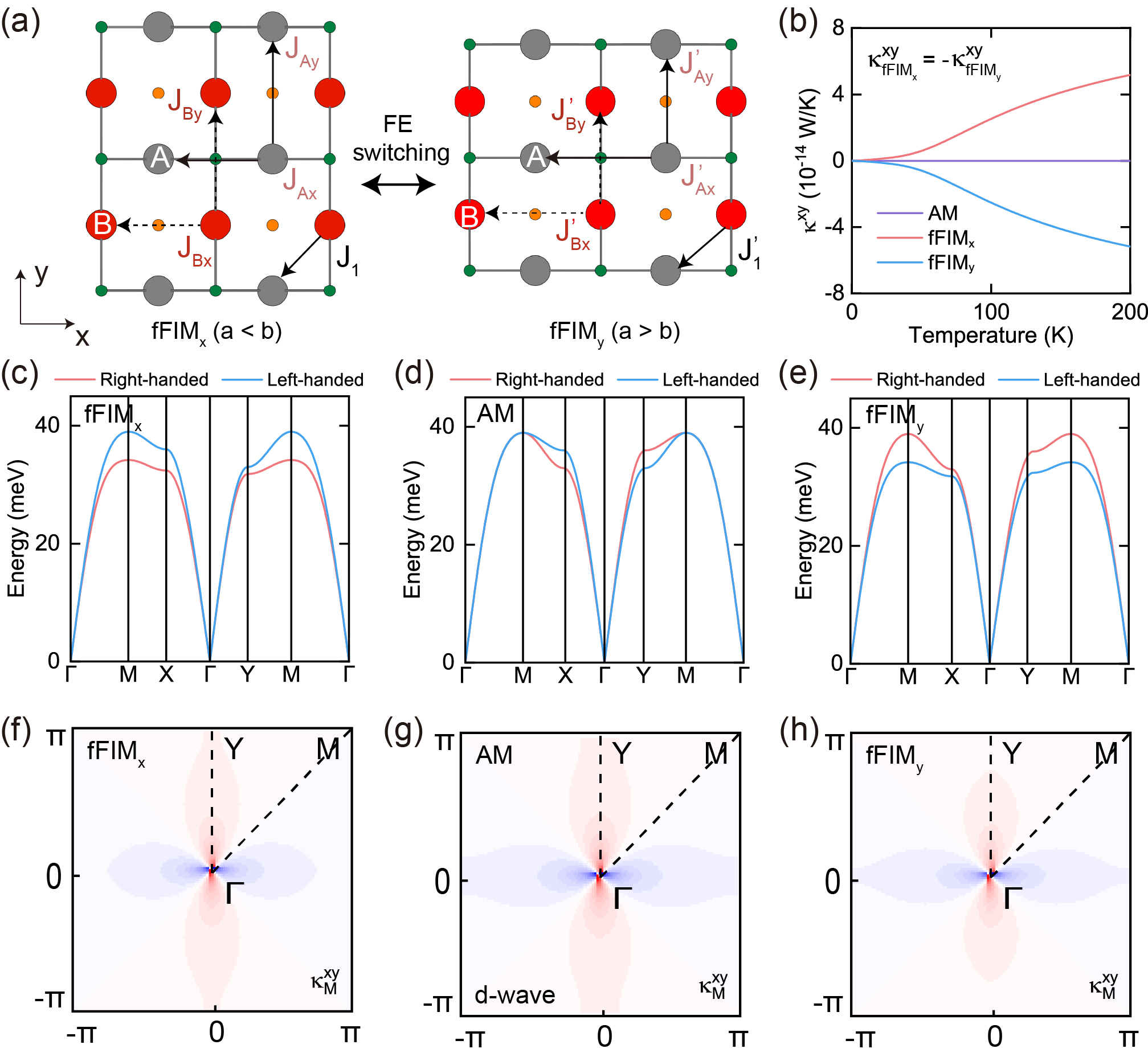}
\caption{ (a) Sketch of the Heisenberg-DM model for the two orthogonal ferroelastic variants. The magnetic sublattices are labeled $A$ and $B$, while $J_1$ and $J_{\alpha\mu}$ $(\alpha=A,B $ and $ \mu=x,y)$ denote the NN and direction-dependent NNN Heisenberg exchange interactions, respectively. The primed parameters refer to the exchange interactions after ferroelastic switching. (b) Temperature dependence magnon thermal Hall conductivity of the Heisenebrg-DM model for the $\mathrm{fFIM}_x$, altermagnet (AM), and $\mathrm{fFIM}_y$ states. (c-e) Handedness-resolved magnonic band structures of the Heisenebrg-DM model for the three magnetic states. Right- and left-handed magnons are denoted by red and blue lines, respectively. (f-h) Momentum-resolved contributions $\kappa_M^{xy}(\mathbf{k})$ to the magnon thermal Hall conductivity of the Heisenebrg-DM model for for the three magnetic states.}
\label{model}
\end{figure}

\textcolor{blue}{\textit{Rationale and model}}---To uncover the symmetry origin of the MTHE reversal under ferroelastic switching, we introduce a Heisenberg-DM model on a two-dimensional rectangular lattice, as illustrated in Fig.~\ref{model}(a). The model Hamiltonian is given by:
\begin{align}
	\begin{split}
		H = &\sum\limits_{(ij)\in N }{J_{ij}}\cdot(\bm{S}_i \cdot \bm{S}_j)+\sum\limits_{\langle ij  \rangle} \bm{D}_{ij} \cdot (\bm{S}_i \times \bm{S}_j) \\
&+A_{ij}^z\sum\limits_{ i }(S_i^z)^2,
	\end{split}
\label{eq}
\end{align}
where $N$ includes both the nearest-neighbor (NN) and next-nearestneighbor (NNN) interactions. $J_{ij}$ is  the Heisenberg exchange coupling between spins $\bm{S_{i/j}}$, whereas $\bm{D_{ij}}$  represents the out-of-plane NN Dzyaloshinskii-Moriya interaction (DMI) vectors, and $A_{ij}^z$ indicate the single-ion anisotropy. For the model calculations presented in Fig.~\ref{model}, we take \(A_z=0\) to focus on the effects of anisotropic exchange and DMI. Based on linear spin-wave theory, the spin operators are expressed in terms of bosonic creation and annihilation operators using the Holstein-Primakoff transformation~\cite{holstein1940field,colpa1978diagonalization}. The expansion is formulated about a collinear 
Néel-type magnetic configuration in which the two magnetic sublattices A and B ($S=3/2$) are polarized along $+\hat{\bm{z}}$ and $-\hat{\bm{z}}$, respectively. Further details of the Heisenberg-DM model are provided in the Supplemental Material~\cite{supp}.

Without loss of generality, we fix the NN parameters as $J_1=5$ meV and $D_1=-0.2$ meV, all the parameters are scaled with the NN Heisenberg exchange parameter $J_1$. As illustrated in Fig.~\ref{model}(a), the relevant spin-space symmetry is governed by the NNN Heisenberg exchange parameters, $J_{Ax}$, $J_{Ay}$, $J_{Bx}$, and $J_{By}$. For $J_{Ax}=J_{By}=-0.2 J_1$ and $J_{Bx}=J_{Ay}=-0.1 J_1$, the resulting magnetic configuration therefore preserves $[C_2 || C_{4z}]$ symmetry. Accordingly, the right- and left-handed magnon branches satisfy $\omega_R^A(\mathbf k)=\omega_L^A(C_{4z}\mathbf k)$, which produces the characteristic $d$-wave momentum-dependent magnonic band splitting of the altermagnetic phase, as shown in Fig.~\ref{model}(d). An in-plane uniaxial lattice perturbation further induces inequivalence among the NNN Heisenberg exchange parameters, thereby breaking the AM phase. Take $J_{Ax}=-0.2J_1$, $J_{Ay}=-0.1J_1$, $J_{Bx}=-0.08J_1$, and $J_{By}=-0.06J_1$ as an example, all symmetries connecting the opposite chiral are broken, and the model enters the fFIM phase, producing a ferromagnet-like global chirality splitting, as shown in Fig.~\ref{model}(c).

\begin{figure}
\centering
\includegraphics[width=8.5cm]{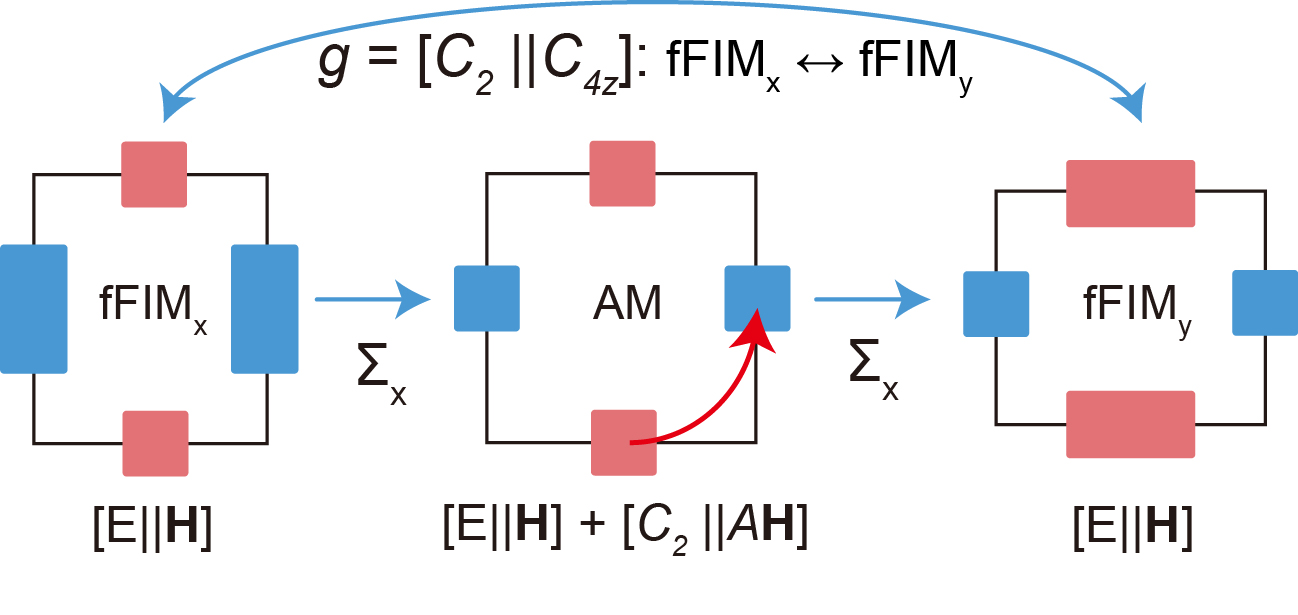}
\caption{Schematic illustration of the magnetic-configuration evolution and accompanying spin-group symmetry reconstruction during ferroelastic switching. The two fFIM states are mapped onto each other by spin-space operation $[C_2\parallel C_{4z}]$, and the corresponding spin groups are indicated below each configuration. Red and blue blocks denote opposite spin sublattices. Here, $C_2$ denotes a twofold spin rotation about an axis perpendicular to the spin axis, $H$ denotes a halving subgroup of the crystallographic Laue group, $A$ denotes a real-space rotation or mirror operation ($C$ or $M$).  }
\label{sch}
\end{figure}

The coupling between ferroelasticity and fFIM phases enables reversible and nonvolatile switching between two orthogonal variants, denoted by $\mathrm{fFIM}_x$ $(a<b)$ and $\mathrm{fFIM}_y$ $(a>b)$, as schematically illustrated in Fig.~\ref{model}(a). Under an $x$-directed tensile strain, the lattice parameter $a$ increases relative to $b$, and the magnetic configuration evolves continuously from $\mathrm{fFIM}_x$, through the AM state at the intermediate configuration, to $\mathrm{fFIM}_y$ state. As shown in Fig.~\ref{sch}, although neither fFIM variant individually possesses a symmetry connecting its opposite-chirality sublattices, the two variants are mapped onto each other by spin-space operation $[C_2||C_{4z}]$, which maps the NNN exchange parameters between the two ferroelastic variants according to $ J'_{Ax}=J_{By}, J'_{Ay}=J_{Bx}, J'_{Bx}=J_{Ay}, J'_{By}=J_{Ax}$. Their magnonic band structures therefore satisfy $\omega_R^{y/x}(\mathbf{k})=\omega_L^{x/y}(C_{4z}\mathbf{k})$, and the chirality splitting obeys $\Delta\omega_{y/x}(\mathbf{k})=-\Delta\omega_{x/y}(C_{4z}\mathbf{k})$, where $\Delta\omega=\omega_R-\omega_L$. Thus, ferroelastic switching rotates the momentum-space magnon spectrum by $90^\circ$ and simultaneously reverses the chirality splitting, as demonstrated by the magnonic band structures of $\mathrm{fFIM}_y$ in Fig.~\ref{model}(e).

We then investigate the ferroelastic control of magnon transport properties, where the thermal Hall conductivity  $\kappa^{xy}$ is given by 
\begin{align}
\begin{split}
 \kappa^{xy} & = -\frac{k^2_BT}{(2\pi)^2\hbar}\sum_{n}\int_{BZ}{c_2(n_B(\varepsilon_n))}{\Omega}_{n}^{xy}({k})d^2\boldsymbol k, 
\end{split}
\end{align}
where $n_B(\varepsilon_n)$ represents the Bose distribution function with $n_B(\varepsilon_n)$ =  $(e^{\frac{\varepsilon_n - \mu}{k_BT}}-1)^{-1}$ and $\mu$ indicating the chemical potential, $c_2(x) = (1+x)\mathrm{ln}^2[(1+x)/x]-\mathrm{ln}^2x-2\mathrm{Li}_2(-x)$, and $\mathrm{Li}_2$ is the polylogarithm function. Consequently, the ferroelastic-switching-induced reconstruction of the magnon spectrum and Berry curvature distribution enables control over both the magnitude and sign of the magnon thermal Hall response. The momentum-resolved contributions to the magnon thermal Hall conductivity are illustrated in Figs.~\ref{model}(f)-(h), where red and blue represent positive and negative contributions, respectively.  In the AM phase, the $[C_2||C_{4z}]$ symmetry connects the opposite-chirality magnon branches and imposes $\mathcal{\kappa}_{\mathrm{A}}^{xy}(\mathbf{k})=-\mathcal{\kappa}_{\mathrm{A}}^{xy}(C_{4z}\mathbf{k})$, giving rise to the $d$-wave momentum-resolved contributions in Fig.~\ref{model}(g) and a vanishing net thermal Hall conductivity. However, the absence of any symmetry relating the opposite-chirality sublattices within either fFIM phase removes this cancellation and permits a finite thermal Hall response.  Owing to the $[C_2\||C_{4z}]$ relation between the two ferroelastic variants, the momentum-resolved response of $\mathrm{fFIM}_y$ becomes the $90^\circ$-rotated and sign-reversed counterpart of that of $\mathrm{fFIM}_x$, as shown in Figs.~\ref{model}(f) and \ref{model}(h). Consequently, their integrated thermal Hall conductivities have equal magnitudes but opposite signs, satisfying $\kappa_{\mathrm{fFIM}_x}^{xy}=-\kappa_{\mathrm{fFIM}_y}^{xy}$, as illustrated in Fig.~\ref{model}(b).

\begin{figure}
\centering
\includegraphics[width=8.5cm]{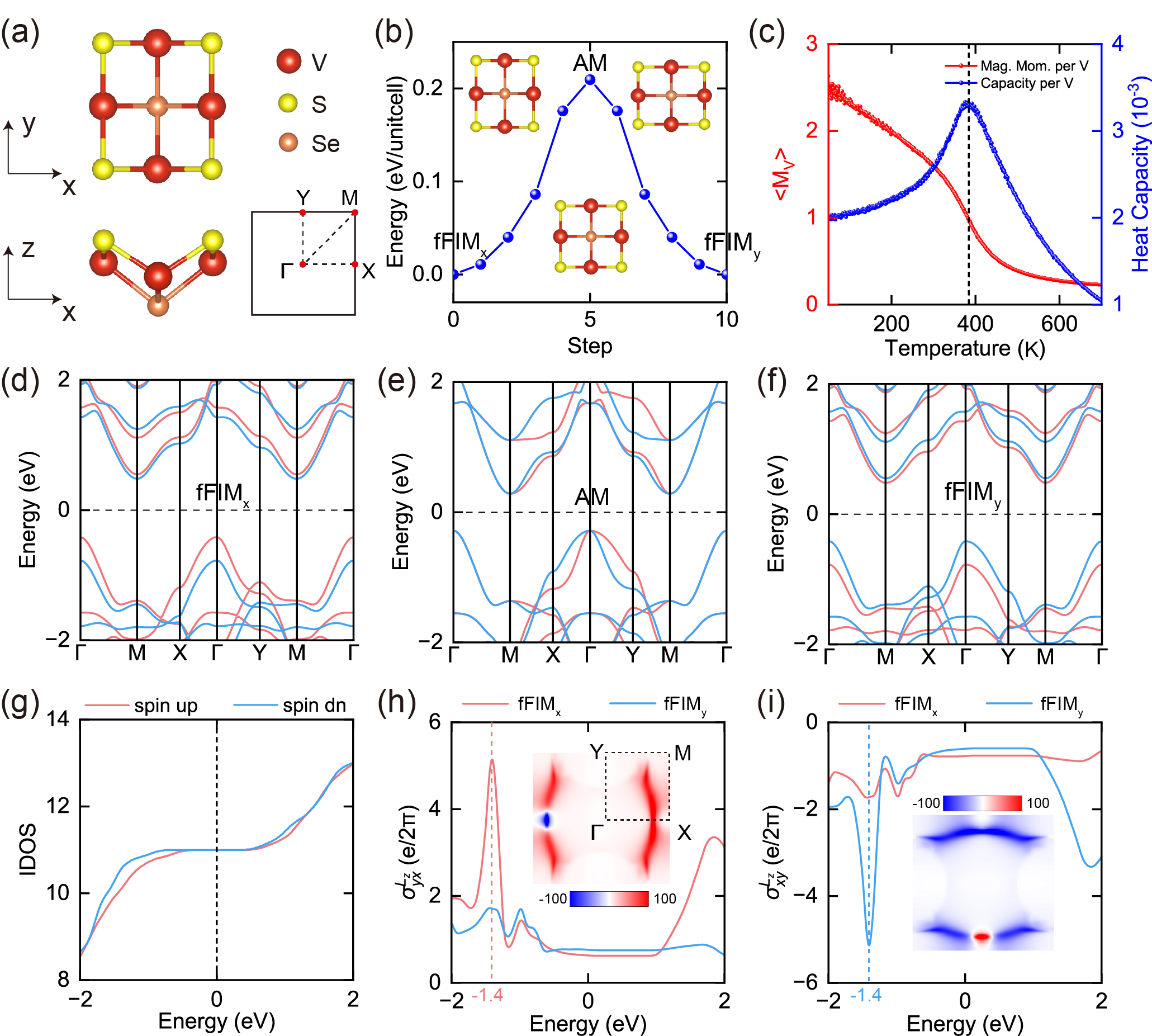}
\caption{(a) Top and side views of the V$_2$SSe monolayer, and the corresponding first Brillouin zone used for the band-structure calculations. (b) Energy pathway for ferroelastic switching as a function of step number in the nudged elastic band calculations. Insets show the corresponding magnetic configurations and crystal structures. (c) Evolution of the average magnetic moment and heat capacity per V atom. (d-f) Spin-resolved electronic band structures of $\mathrm{fFIM}_x$, AM, and $\mathrm{fFIM}_y$, respectively. (g) Spin-resolved integrated densities of states of $\mathrm{fFIM}_x$ state. (h,i) Energy dependence of the orbital Hall conductivities for longitudinal driving fields along the $x$ and $y$ directions, respectively. The insets display the corresponding momentum-resolved orbital Berry curvatures $\Omega_{yx}^{L_z}(\mathbf{k})$ for $\mathrm{fFIM}_x$ and $\Omega_{xy}^{L_z}(\mathbf{k})$ for $\mathrm{fFIM}_y$ at $E_F=-1.4$ eV. }
\label{mat}
\end{figure}

\textcolor{blue}{\textit{Material Realization}}---Guided by the model analysis, we select monolayer V$_2$SSe as a candidate system, in which ferroelasticity, fully compensated magnetic order, and symmetry-allowed Dzyaloshinskii-Moriya interactions provide the essential ingredients for switchable Berry-curvature-driven magnon transport. As illustrated in Fig.~\ref{mat}(a), monolayer V$_2$SSe is a two-dimensional collinear antiferromagnet with a noncentrosymmetric rectangular lattice, belonging to the magnetic space group $Pm^{\prime}m^{\prime}2$. The optimized lattice constants are $a=4.12$~\AA{} and $b=4.77$~\AA{}. The pronounced in-plane structural anisotropy makes V$_2$SSe a natural candidate for realizing the ferroelastic magnetic switching predicted by our model. The minimum-energy pathway connecting the two orthogonal ferroelastic variants is further examined using the nudged elastic band (NEB) method. As shown in Fig.~\ref{mat}(b), $\mathrm{fFIM}_x$ and $\mathrm{fFIM}_y$ are degenerate lowest-energy configurations. The resulting switching barrier is approximately $\Delta E_{\mathrm b}=$ 0.22 eV/unitcell, which is comparable to that reported for representative two-dimensional ferroelastic materials such as monolayer $1T^\prime$-WTe$_2$~\cite{li2016ferroelasticity}. The absence of imaginary modes throughout the Brillouin zone in the calculated phonon spectrum further confirms the fFIM phase of V$_2$SSe monolayer is dynamically stable~\cite{supp}.

The magnetic exchange parameters of monolayer V$_2$SSe are obtained through the Green's-function method based on the magnetic force theorem~\cite{he2021tb2j}, and the resulting exchange parameters are listed in Supplemental Material~\cite{supp}. The dominant NN exchange interaction $J_1= 22.16$ meV, favors antiparallel alignment between neighboring V atoms, while the magnetic anisotropy $A_z=-0.058$ meV stabilizes their orientation along the out-of-plane direction. Figures~\ref{mat}(d-f) present the electronic band structures of the three magnetic phases with spin-orbit coupling. In both $\mathrm{fFIM}_x$ and $\mathrm{fFIM}_y$ states, the absence of the spin-space group operation connecting the opposite-spin sublattices lifts the spin degeneracy, resulting in globally spin-split electronic bands. Moreover, the electronic band structures of the two fFIM variants are related by the operation $[C_2 || C_{4z}]$, exhibiting the $90^\circ$-rotated momentum-dependent spin splitting and spin-reversed correspondence. Nevertheless, the spin-resolved integrated densities of states (IDOS) satisfy $N_{\uparrow}=N_{\downarrow}$, resulting in a vanishing net magnetization and demonstrated the filling-enforced fFIM phase, as shown in Fig.~\ref{mat}(g) and Fig. S3~\cite{supp}. While the intermediate AM phase resulting in $d$-wave momentum-dependent spin splitting and degenerate spin-resolved IDOS, as shown in Fig.~\ref{mat}(e) and Fig. S3~\cite{supp}.

\begin{figure}
\centering
\includegraphics[width=8.5cm]{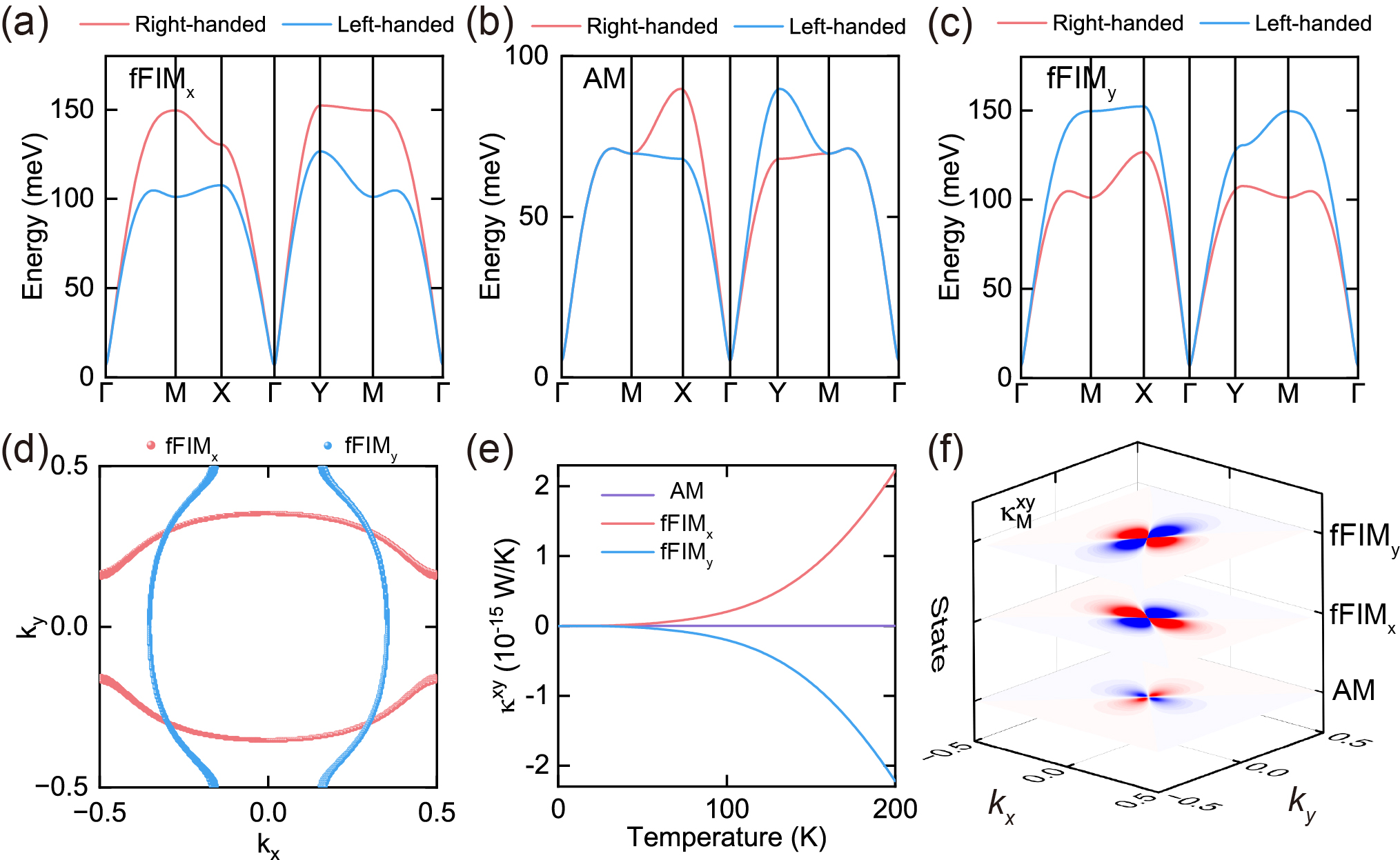}
\caption{(a-c) Magnonic band structures of V$_2$SSe monolayer for the $\mathrm{fFIM}_x$, AM, and $\mathrm{fFIM}_y$ states. Right- and left-handed magnons are denoted by red and blue lines, respectively. (d) Momentum-space distribution of magnon band structures near $E=130$ meV for the $\mathrm{fFIM}_x$ (right-handed) and $\mathrm{fFIM}_y$ (left-handed) states. The plotted points satisfy $\left|\varepsilon-130\ \mathrm{meV}\right|\leq0.5\ \mathrm{meV}$. (e) Temperature dependence magnon thermal Hall conductivity V$_2$SSe monolayer for the three magnetic states. (f)  The distribution of momentum resolved thermal Hall conductivity $\kappa_M^{xy}(\mathbf{k})$  of V$_2$SSe monolayer at 150 K in whole Brillouin zone. }
\label{magnon}
\end{figure}


The ferroelastic reconstruction of the electronic bands is also reflected in the orbital Hall response. Within linear-response theory, the orbital Hall conductivity (OHC) is given by~\cite{chen2024topology}
\begin{align}
\begin{split}
\sigma_{ij}^{L_z}=-e\int_{\mathrm{BZ}}\frac{d^{2}\mathbf{k}}{(2\pi)^{2}}\sum_{n}f_{n}(\mathbf{k})\Omega_{n,ij}^{L_z}(\mathbf{k}),
\end{split}
\end{align}
where $f_n(\mathbf{k})$ and $\Omega_{n,ij}^{L_z}(\mathbf{k})$ denote the Fermi-Dirac distribution and orbital Berry curvature, respectively~\cite{supp}. Figures~\ref{mat}(h) and \ref{mat}(i) show $\sigma_{yx}^{L_z}$ and $\sigma_{xy}^{L_z}$ for electric fields applied along the $x$ and $y$ directions, respectively. Consistent with the $[C_2 || C_{4z}]$ relation, the two fFIM variants satisfy
$\sigma_{yx}^{L_z}(\mathrm{fFIM}_x)=-\sigma_{xy}^{L_z}(\mathrm{fFIM}_y)$. The insets reveal $90^\circ$-rotated and sign-reversed orbital Berry curvature hot spots near the $X$ and $Y$ points at $E-E_F=-1.4$ eV, corresponding to the OHC extrema. Despite the topologically trivial electronic bands, a finite OHC $\sigma_{yx}^{L_z}$ persists at the intrinsic Fermi level with magnitudes of approximately $0.7 e/2\pi$ ($\mathrm{fFIM}_x$) and $0.6 e/2\pi$ ($\mathrm{fFIM}_y$), whereas the spin Hall conductivity remains negligible, as shown in Fig.~\ref{mat}(h). The resulting orbital-dominated Hall response provides a comparatively clean setting for identifying orbital transport.

\begin{figure}
\centering
\includegraphics[width=8.5cm]{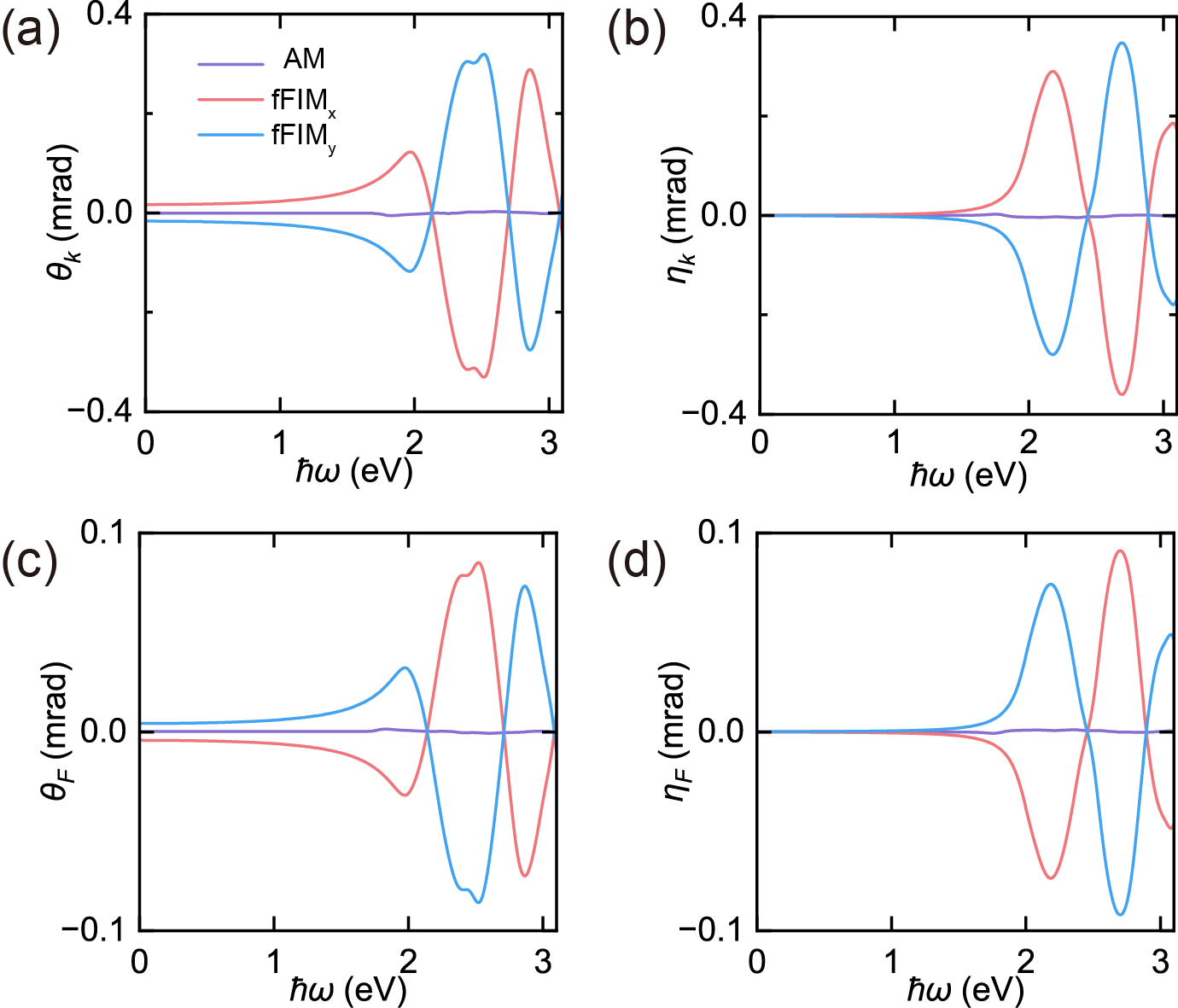}
\caption{Magneto-optical fingerprints of the ferroelastic magnetic states. Photon-energy dependence of (a) the Kerr rotation angle $\theta_K$, (b) the Kerr ellipticity $\eta_K$, (c) the Faraday rotation angle $\theta_F$, and (d) the Faraday ellipticity $\eta_F$ for the AM, $\mathrm{fFIM}_x$, and $\mathrm{fFIM}_y$ phases. }
\label{moke}
\end{figure}

Having confirmed the magnetic configuration, we now focus on the magnonic properties of monolayer V$_2$SSe. As shown in Figs.~\ref{magnon}(a)-\ref{magnon}(c), the $\mathrm{fFIM}_x$ and $\mathrm{fFIM}_y$ phases exhibit ferromagnet-like global chirality splitting, with the energy separation between the right- and left-handed magnon branches reaching approximately $48$ meV at the $M$ point, whereas the intermediate AM phase retains the characteristic $d$-wave momentum-dependent splitting. The interchange of the right- and left-handed branches under ferroelastic switching, together with the accompanying $90^\circ$ rotation of the constant-energy contours constructed within $130\pm0.5$ meV [Fig.~\ref{magnon}(d)], reflects the symmetry relation between the two ferroelastic variants established by the Heisenberg-DM model. Monte Carlo simulations yield a Néel temperature of approximately $380$ K, as shown in Fig.~\ref{mat}(c). To remain sufficiently below the critical regime, where linear spin-wave theory becomes unreliable, the thermal Hall calculations are restricted to $T\leq200$ K. As shown in Fig.~\ref{magnon}(e), $\kappa_{\mathrm{fFIM}_x}^{xy}=-\kappa_{\mathrm{fFIM}_y}^{xy}=2.2\times10^{-15}$ W/K at $200$ K, whereas the net response of the AM phase vanishes. The momentum-resolved distributions in Fig.~\ref{magnon}(f) further reveal the sign-reversed thermal Hall contributions. Taken together with the electronic band reconstruction in Figs.~\ref{mat}(d)-\ref{mat}(f), our results establish dual ferroelastic control over the electronic and magnonic spectra and their associated transverse responses.

\textcolor{blue}{\textit{Magneto-optical fingerprints}}---Finally, we explore magneto-optical response as a contact-free approach for identifying the three ferroelastic magnetic states. Defining the Kerr/Faraday angle as $\theta_{K/F}(\omega)+i\eta_{K/F}(\omega)$, the expressions for normally incident $x$-polarized light are given by~\cite{kim2007determination,valdes2012terahertz}
\begin{align}
\begin{split}
\theta_K+i\eta_K&=\frac{2Z_0d\sigma_{xy}}{\left[1-n_s-Z_0d\sigma_{xx}\right]\left[1+n_s+Z_0d\sigma_{yy}\right]}\\
\theta_F+i\eta_F&=\left(\frac{\sigma_{xy}}{\sigma_{yy}}\right)\left[1+\frac{n_s+1}{Z_0d\sigma_{yy}}\right]^{-1},
\end{split}
\end{align}
where $\theta$ and $\eta$ denote the rotation angle and ellipticity, respectively. Here, $Z_0\simeq377~\Omega$ is the vacuum impedance, $d$ is the effective thickness of the monolayer, and $n_s\simeq1.5$ is the refractive index adopted for the substrate~\cite{PhysRevLett.134.116703}. The quantities $\sigma_{xx}$ and $\sigma_{yy}$ are the longitudinal optical conductivities, while $\sigma_{xy}$ denotes the transverse component. As shown in Figs.~\ref{moke}(a) and \ref{moke}(b), the Kerr response of the AM phase remains nearly zero, whereas the two fFIM phases exhibit pronounced spectra with magnitudes reaching approximately $0.33$ mrad around $\hbar\omega=2.5$ eV. 

The $[C_2 || C_{4z}]$ relation between the two fFIM phases constrains their optical conductivity tensors as
\begin{align}
\boldsymbol{\sigma}^{(y)}(\omega)
&=
R_{4z}\left[\boldsymbol{\sigma}^{(x)}\right]^{T}R_{4z}^{T}=
\begin{pmatrix}
\sigma_{yy}^{(x)} & -\sigma_{xy}^{(x)} \\
-\sigma_{yx}^{(x)} & \sigma_{xx}^{(x)}
\end{pmatrix},
\end{align}
where $R_{4z}$ represents the $C_{4z}$ rotation and the superscripts $x$ and $y$ label the $\mathrm{fFIM}_x$ and $\mathrm{fFIM}_y$ phases, respectively. Ferroelastic switching therefore reverses the optical Hall component while exchanging $\sigma_{xx}$ and $\sigma_{yy}$. Their anisotropy $\sigma_{xx} \neq \sigma_{yy}$ produce sign-contrasting Kerr spectra with unequal magnitudes for a fixed incident polarization, as shown in Figs.~\ref{moke}(a) and \ref{moke}(b).  In the regime  $|Z_0d\sigma_{xx}|,|Z_0d\sigma_{yy}|\ll 1+n_s$, the longitudinal-conductivity terms can be neglected to leading order, giving $\Phi_K\simeq2Z_0d\sigma_{xy}/(1-n_s^2)$ and $\Phi_F\simeq Z_0d\sigma_{xy}/(1+n_s)$.  Consequently, $\Phi_F\simeq(1-n_s)\Phi_K/2=-\Phi_K/4$ for $n_s=1.5$, consistent with Figs.~\ref{moke}(c) and \ref{moke}(d). These state-dependent magneto-optical spectra provide contact-free optical fingerprints of the ferroelastic magnetic states.

\textcolor{blue}{\textit{Conclusions}}---In summary, we have uncovered a symmetry mechanism for ferroelastic reversal of the magnon thermal Hall effect in fully compensated ferrimagnets. Our model analysis shows that ferroelastic switching reorients the anisotropic magnon spectrum, interchanges opposite-chirality branches, and reverses the associated Berry-curvature-driven thermal Hall response without generating a net magnetization. First-principles calculations establish monolayer V$_2$SSe as a promising material realization of this mechanism. The accompanying orbital Hall and magneto-optical responses provide complementary transport and optical fingerprints for distinguishing the switched magnetic states. These findings establish ferroelasticity as an effective route to reversible and nonvolatile control of transverse magnon heat transport in zero-net-magnetization systems.


M. Y. acknowledges financial support from Hong Kong Research Grants Council (project number: 15307625) and The Hong Kong Polytechnic University (project number: P0048122, P0051269 and P0051270). C.N. acknowledges financial support from the Shandong Provincial Science Foundation for Excellent Young Scholars (Grant No. ZR2023YQ001) and Taishan Scholar Program of Shandong Province.


%

\end{document}